\documentstyle[a4,11pt]{article}
{\textwidth 165mm
\textheight 232mm}

\begin{document}

\title{Relativistic particle interaction with a weak electromagnetic field}

\author{L.~F.~Blazhyjevskii,\,\,\,\, A.~Y.~Marko\\
{\it Ivan Franko Lviv National University, Chair of Theoretical Physics}\\
{\it 12 Dragomanov Str., UA 79005, Lviv, Ukraine}\\
{\it E-mail: ya@sotka.ktf.franko.lviv.ua}}
\date{Received}
\maketitle

\begin{abstract}

Schrodinger equation with two-component wave function which 
describes a relativistic spin 1/2 particle in a weak electromagnetic field is 
obtained. In the same approximation Schrodinger equation with traditional
norm condition and one-component wave function for a spinless particle 
is obtained as well. To construct it Foldy-Wouthuysen procedure with the 
electron charge value as the small parameter is used.  
\end{abstract}

\section{Introduction}

Attempts to describe the electron which moves with the speed close to the
speed of light by two-component wave function appeared after the creation of the 
relativistic quantum mechanics. It was connected with the fact that in the
nonrelativistic case the electron is described by two-component wave function.
However, in 1932, Dirac had shown that the relativistic electron is described
by four-component wave function which is the solution of the Dirac equation.
Trying to find Pauli equation as the limit case of Dirac equation Foldy and
Wouthuysen \cite{Foldy} supposed for it the unitary transformation method. 
This method allowed describing the positive energy states of the free particle by 
the "large" components of the four-component spinors 
and negative energy states by the "small" components. 
It means that the positive energy state is described by 
the spinor which has third and fourth components equal to zero and the 
negative energy state is described by spinor with the first and 
second components equal to zero. In the case of the presence external 
electromagnetic field such separation had been done approximately with 1/c as
the small parameter. In this way Schrodinger equation with two-component
wave function was received which is suitable in the weakly relativistic case. 
Berestetskii and Landau received this equation by other method that differs 
from Foldy-Wouthuysen procedure \cite{Berestetskii}. 

However the weakly relativistic equation is not good for the particle which 
moves with the speed very close to the speed of the light. Had used 
Foldy-Wouthuysen procedure the attempt to describe relativistic particle 
by two-component wave function in the presence of a weak external 
electromagnetic field had been done in 1962 \cite{Blount}. But in the 
\cite{Blount} higher derivatives with respect to potentials 
of the electromagnetic field were neglected. In 1995 \cite{Silenko}
Landau-Berestetskii method was used to separate the "large" and the "small"
components in the Dirac equation for the spin 1/2 particle in the weak 
field approximation. But as for us this solution of the problem is not 
quite apprehensible.

In this paper Foldy-Wouthuysen procedure is used to construct Schrodinger
equation with two-component wave function for relativistic electron in a weak
external electromagnetic field. Schrodinger equation for a relativistic 
spinless particle in a weak external electromagnetic is also obtained.
As in \cite{Blount} and \cite{deGroot} we take into account the contributions 
of the first order by small parameter $e$ but we 
take also into account higher derivatives with respect to the field potentials. 
In the special case when we neglect higher orders derivatives we obtain
result \cite{Blount}, \cite{deGroot}.
In \cite{Silenko} higher orders derivatives is taken into account also, but
our method and result differs from method and result \cite{Silenko}.

\section{Relativistic spin-1/2 particle interaction with a weak 
         electromagnetic field}

A free relativistic particle  with spin 1/2 is described by stationary Dirac
equation          
   \begin{equation}
   \label{m1}
    E\psi=\hat H_D\psi;
   \end{equation}
where $\hat H_D$ - Dirac Hamiltonian
   \begin{equation}
   \hat H_D=c\vec{\alpha}\hat{\bf p}+\beta mc^2.
   \end{equation}
$\vec{\alpha}$ and $\beta$ - 4$\times$4 matrix which could be written in the
block form
   \begin{equation} \vec\alpha=\left(\begin{array}{ccc}0&\vec\sigma\\ 
   \vec\sigma&0\\ \end{array} \right);\,\,\,\,
   \beta=\left(\begin{array}{ccc}1&0\\0&-1\\ 
   \end{array} \right);
   \end{equation}
here the sign "$1$" in the matrix $\beta$ means $2\times2$ unit matrix,  
and $\vec{\sigma}$ - $2\times2$ Pauli matrix
   \begin{equation} \sigma_x=\left(\begin{array}{ccc}0&1\\ 
                              1&0\\ \end{array} \right);\,\,\,\,
   \sigma_y=\left(\begin{array}{ccc}0&-i\\ 
                  i&0\\ \end{array} \right);\,\,\,\,
   \sigma_z=\left(\begin{array}{ccc}1&0\\ 
                   0&-1\\ \end{array} \right).
   \end{equation} 
 
Foldy and Wouthuysen showed how one could find such representation
where the wave function of a free particle with respect to positive energy 
state is spinor with the third and the fourth components equal to 
zero. Spinor with the first and the second components equal to zero respond
to the negative energy state. In Foldy-Wouthuysen representation the first 
two components of the wave function are called "large" components 
and the second two components are called "small" components. 
Beside that, Dirac Hamiltonian has the diagonal
form in Foldy-Wouthuysen representation. 
It means that Dirac equation can be written as the
system of two equations. The first equation of this system contains only the 
"large" components and the second equation contains only the "small" components. 
Therefore one can separate "large" and "small" components.
The unitary operator makes transmission to Foldy-Wouthuysen representation
$U_0(\hat{\bf p}):\Psi=U_0(\hat{\bf p})\psi$
   \begin{equation}
   U_0(\hat{\bf p})U_0^+(\hat{\bf p})=U_0^+(\hat{\bf p})U_0(\hat{\bf p})=1;
   \end{equation}
   \begin{equation}
   U_0(\hat{\bf p})=
   \frac{1}{\sqrt{2\varepsilon(\hat{\bf p})(\varepsilon(\hat{\bf p})+mc^2)}}
   \left(\varepsilon(\hat{\bf p})+mc^2+\beta c\vec{\alpha}{\bf\hat p}\right). 
   \end{equation}
In this way Dirac equation for the free particle will be the next
   \begin{equation}
   E\Psi=U_0(\hat{\bf p})\hat H_DU_0^+(\hat{\bf p})\Psi=
   \beta\varepsilon(\hat{\bf p})\Psi;
   \end{equation}
where $\varepsilon(\hat{\bf p})=\sqrt{m^2c^4+c^2\hat{\bf p}^2}$ - the 
relativistic energy operator of the free particle.

The equation for Dirac particle in the stationary electromagnetic field
is obtained from equation (\ref{m1}) by well known substitution
$E\rightarrow E-e\Phi$,
$\hat{\bf p}\rightarrow\vec\pi=\hat{\bf p}-e/\!c\,{\bf A}$ 
   \begin{equation} \label{m2}
   E\psi=\left[c\vec{\alpha}\vec{\pi}+\beta mc^2+e\Phi\right]\psi.
   \end{equation}
In the presence of the external electromagnetic field it is impossible to
separate "large" and the "small" components in the equation (\ref{m2}) 
by means of Foldy-Wouthuysen procedure exactly. 
It is possible to do it approximately using a perturbation theory 
with a small parameter. Foldy and Wouthuysen supposed the 
inverted value of the light speed $1/c$ as the small parameter. 
It means that one
considers the case when the particle energy is not very large. 
In our work the value of the electron charge $e$ is the small
parameter. This value characterises the electron interaction with the 
electromagnetic field. Therefore we consider the case of the weak particle
interaction with the field and the arbitrary particle energy. Aiming to get the 
diagonal Hamiltonian let us do the unitary transformation in the equation 
(\ref{m2}) by means of the operator 
$U(\vec\pi):\Psi=U(\vec\pi)\psi$.
   \begin{equation}
   U(\vec\pi)=
   \frac{1}{\sqrt{2\varepsilon(\vec\alpha\vec\pi)
   (\varepsilon(\vec\alpha\vec\pi)+mc^2)}}
   \left(\varepsilon(\vec\alpha\vec\pi)+mc^2+
   \beta c\vec\alpha\vec\pi\right). 
   \end{equation}
The transformed equation (\ref{m2}) becomes  
   \begin{equation}  
   \label{m3}
   E\Psi=
   \left[\beta\varepsilon(\vec\Sigma\vec\pi)+e\hat\Phi_d+e\hat\Phi_{nd}
   \right]\Psi;
   \end{equation}
where
   \begin{equation}
   \label{Phi_d}
   \hat\Phi_d=\frac{1}{2}
   \left\{U_0(\hat{\bf p})\Phi U_0^+(\hat{\bf p})+
   U_0^+(\hat{\bf p})\Phi U_0(\hat{\bf p})\right\}\,;
   \end{equation}
   \begin{equation}
   \hat\Phi_{nd}=\frac{1}{2}
   \left\{U_0(\hat{\bf p})\Phi U_0^+(\hat{\bf p})-
   U_0^+(\hat{\bf p})\Phi U_0(\hat{\bf p})\right\}\,;
   \end{equation}
and $\vec\Sigma$ is $4\times4$ matrix
   \begin{equation}
   \vec\Sigma=\left(\begin{array}{ccc}\vec\sigma&0\\ 
   0&\vec\sigma\\ \end{array} \right).
   \end{equation}    
The obtained Hamiltonian is diagonal in the zeroth order in the constant of 
the interaction. The first order contains the diagonal part $\hat\Phi_d$ and 
the nondiagonal part $\hat\Phi_{nd}$. In order to exclude the nondiagonal 
part it's necessary to manage a unitary transformation ones again.
After that we would obtain the diagonal Hamiltonian in the zeroth and in the 
first orders. We also would obtain the terms of the second order in the
small parameter. Those terms are small and we do not take them into account.
In this way the "large" and the "small" components can be separated in the 
equation for relativistic spin 1/2 particle in the weak electromagnetic field.
But the separation of components can be done by other method. In order to do 
it let us write operators $\hat\Phi_d$ and $\hat\Phi_{nd}$ in the next form
   \begin{equation}
   \label{fi_d}
   \hat\Phi_d=\left(\begin{array}{ccc}\Phi'&0\\ 
   0&\Phi'\\ \end{array} \right)\,;\,\,\,\,\,\,
   \hat\Phi_{nd}=\left(\begin{array}{ccc}0&\Phi''\\ 
   -\Phi''&0\\ \end{array} \right)\,;
   \end{equation}
here   
   \begin{equation}
   \label{fi'}
   \Phi'=
   \frac{(2\varepsilon(\hat{\bf p}))^{-1/2}}
   {(\varepsilon(\hat{\bf p})+mc^2)^{\frac{1}{2}}}
   \left((\varepsilon(\hat{\bf p})+mc^2)\Phi(\varepsilon(\hat{\bf p})+mc^2)+
   c^2\vec\sigma{\bf\hat p}\Phi\vec\sigma{\bf\hat p}\right)
   \frac{(2\varepsilon(\hat{\bf p}))^{-1/2}}
   {(\varepsilon(\hat{\bf p})+mc^2)^{\frac{1}{2}}}\,;
   \end{equation}
   \begin{equation}
   \Phi''=
   \frac{c(2\varepsilon(\hat{\bf p}))^{-1/2}}
   {(\varepsilon(\hat{\bf p})+mc^2)^{\frac{1}{2}}}
   \left(\vec\sigma{\bf\hat p}\Phi(\varepsilon(\hat{\bf p})+mc^2)+
   (\varepsilon(\hat{\bf p})+mc^2)\Phi\vec\sigma{\bf\hat p}\right)
   \frac{(2\varepsilon(\hat{\bf p}))^{-1/2}}
   {(\varepsilon(\hat{\bf p})+mc^2)^{\frac{1}{2}}}\,.
   \end{equation}
Taking into the account that
   \begin{equation}
   \Psi=\left(\begin{array}{cc}\phi\\ 
   \chi\\ \end{array}\right)\,;
   \end{equation}
we can write the equation (\ref{m3}) in the next form
   \begin{equation}
   \label{m5}
   E\phi=\varepsilon(\vec\sigma\vec\pi)\phi+e\Phi'\phi+e\Phi''\chi\,;
   \end{equation}
   \begin{equation} 
   \label{m4}
   E\chi=-\varepsilon(\vec\sigma\vec\pi)\chi+e\Phi'\chi-e\Phi''\phi\,.
   \end{equation} 
Let us express the function $\chi$ from the equation (\ref{m4}) in the case
of the positive energy, and the function $\phi$ from the equation
(\ref{m5}) in the case of the negative energy
   \begin{eqnarray}
   \label{chi}
   \chi=-(E+\varepsilon(\vec\sigma\vec\pi)-e\Phi')^{-1}e\Phi''\phi\,,\\
   \label{phi}
   \phi=(E-\varepsilon(\vec\sigma\vec\pi)-e\Phi')^{-1}e\Phi''\chi\,;
   \end{eqnarray}
i.~e. $\chi\sim e\phi$ when $E>0$, and $\phi\sim e\chi$ when $E<0$. 
After putting (\ref{chi}) into the equation (\ref{m5}) and (\ref{phi}) into 
the equation (\ref{m4}) and neglecting the terms of the second orders we get
the approximate Dirac equation in Foldy-Wouthuysen representation
   \begin{equation}  
 \label{FW_for_Dirac_eq}
   E\left(\begin{array}{cc}\phi\\ \chi\\ \end{array}\right)=
   \left[\beta\varepsilon(\vec\Sigma\vec\pi)+e\hat\Phi_d\right]
   \left(\begin{array}{cc}\phi\\ \chi\\ \end{array}\right).
   \end{equation}
Here the function $\phi$ responds to the positive energy state and the 
function $\chi$ responds to the negative energy state.

If we neglect the contribution of the second order then the norm 
condition of the wave function $\Psi$
   \begin{equation}
   \int\!\Psi^+\Psi dv=\int(\phi^+\phi+\chi^+\chi)dv=1;
   \end{equation}
can be written using only the function $\phi$ in the case $(E>0)$  
   \begin{equation}
   \label{m10}
   \int\!\phi^+\phi\,dv=1\,.
   \end{equation}
Therefore we can write two-component Schrodinger equation for the relativistic
electron in the weak external electromagnetic field  
   \begin{equation}
   \label{m6'}
   E\phi=\hat H\phi\,;
   \end{equation}
where   
   \begin{equation}
   \label{m6}
   \hat H=\varepsilon\left(\vec\sigma\vec\pi\right)+e\Phi'.
   \end{equation}

From ($\ref {m6}$) It is easy to get Pauli spin and weakly relativistic 
Hamiltonians \cite{Berestetskii}. For that one should expand (\ref{m6}) 
in series $1/c$ and take the terms of the zeroth and the first orders. 
Therefore result (\ref{m6'}) and (\ref{m6}) is a generalisations of Pauli spin
equation and weakly relativistic equation.

It is also possible to manage Weil transformation in the (\ref{FW_for_Dirac_eq}).
Weil transformation for the operator $\hat A$ is function $a(q,p), 
(a\leftrightarrow\hat A)$ which is determined by the expression \cite{deGroot}
   $$a(q,p)=\int\!du\,e^{\frac{i}{\hbar}\bf qu}\langle{\bf p}+\frac{1}{2}{\bf u}|
   \hat A|{\bf p}-\frac{1}{2}{\bf u}\rangle\,;$$
here $|{\bf p}-\frac{1}{2}{\bf u}\rangle$ - is eigenvector of the impulse operator
which responds to eigenvalue ${\bf p}-\frac{1}{2}{\bf u}$. If the operator
$\hat A$ is the function which depends only on $\bf\hat p$ or only on 
$\bf\hat r$ then Weil transformation is the same function which depends on 
$\bf p$ or $\bf q$ respondly. Weil transformation of the multiplication 
of two operator responds expression which is determined by Weil 
transformation of every operator. In order to find Weil transformation 
for operator $\Phi_d$ which is determined by (\ref{Phi_d}) one should know
Weil transformation which responds to multiplication of three operators.
According to \cite{deGroot} the formula 
  \begin{eqnarray}
  \nonumber\hat A\hat B\hat C\leftrightarrow\nonumber\exp\left\{\frac{i\hbar}{2}
  \left(\frac{\partial^{(b)}}{\partial {\bf q}}
  \frac{\partial^{(c)}}{\partial {\bf p}}-
  \frac{\partial^{(b)}}{\partial{\bf q}}\frac{\partial^{(a)}}{\partial{\bf p}}
  \right)\right\}
  \sum\limits_{\mu,\nu}a_{k,\mu}b_{\mu,\nu}c_{\nu,l}
  \end{eqnarray}
is true for the multiplication of three operators $\hat A(\hat{\bf p})$, 
$\hat B({\bf r})$, $\hat C(\hat{\bf p})$ which act on spinors and have Weil 
transformation $a_{k,\mu}({\bf p})$, $b_{\mu,\nu}({\bf r})$, 
$c_{\nu,l}({\bf p})$. Let us write Weil transformation for the Hamiltonian 
(\ref{FW_for_Dirac_eq}). The transmission formulas to this representation 
contain higher derivatives with respect to the potentials of the 
electromagnetic field. If they are neglected then Blount result
is obtained \cite{Blount}, \cite{deGroot}
  \begin{equation}
  \hat H \leftrightarrow\beta\varepsilon(\vec\pi)+e\Phi-
  \mu_B\frac{mc^2}{\varepsilon({\bf p})}\beta\vec\Sigma{\bf B}-
  \mu_B\frac{mc^3}{\varepsilon({\bf p})(\varepsilon({\bf p})+mc^2)}
  \left[{\bf p}\times\vec\Sigma\right]{\bf E}
  \end{equation}
where $\mu_b$ - Bor magneton. 
  
Let us consider the operator $A$ of Dirac theory in Foldy-Wouthuysen
representation for the positive energy states. In Dirac theory operator $A$ 
act on four component wave function. Let us respond it operators which act only
on "large" components. If operator $UAU^+$ is diagonal the matrix element 
$\langle\Psi'|UAU^+|\Psi''\rangle$ can be written as
   \begin{equation}
   \label{D_op}
   \langle\phi'|UAU^+|\phi''\rangle+\langle\chi'|UAU^+|\chi''\rangle
   \end{equation} 
here
   $$|\phi\rangle=\left(\begin{array}{c}\phi\\0\end{array}\right)\,;\,\,\,\,\,\,
   |\chi\rangle=\left(\begin{array}{c}0\\\chi\end{array}\right)\,.$$   
The second term in the (\ref{D_op}) is of the second order in small parameter 
$e$ and can be dropped down. Therefore the operator of the Dirac theory $A$ 
should be replaced by the projection of the operator $UAU^+$ on the subspace 
of "large" components. If the operator $UAU^+$ is nondiagonal then the matrix 
element $\langle\Psi'|UAU^+|\Psi''\rangle$ can be written as
   \begin{equation}
   \label{NDML}
   \langle\phi'|UAU^+|\chi''\rangle+\langle\chi'|UAU^+|\phi''\rangle\,.
   \end{equation} 
Here we can neglect nothing. But from expression (\ref{chi}) we can find
   \begin{equation}
   \label{CHI_}
   |\chi\rangle=\rho_1\frac{e}{E+\varepsilon({\bf\hat p})}\Phi''|\phi\rangle\,;
   \end{equation}    
where $\rho_1$ is $4\times4$ matrix
   \begin{equation}
   \rho_1=\left(\begin{array}{cc}0&1\\1&0\end{array}\right)\,.
   \end{equation}   
Therefore, from expressions (\ref{CHI_}) (\ref{NDML}) the operator of Dirac
theory $A$ should be replaced by the projection of the operator
   \begin{equation}
   \left(\rho_1\frac{e}{E+\varepsilon({\bf\hat p})}\Phi''\right)^+UAU^++
   UAU^+\rho_1\frac{e}{E+\varepsilon({\bf\hat p})}\Phi''\,;
   \end{equation}
on the subspace of "large" components. In this place the remark must be done.
The operators $(E+\varepsilon({\bf p}))^{-1}$ and $\Phi''$ in the expressions
(\ref{NDML}) and (\ref{CHI_}) should be permuted. 
It can be done when we use Fourier transformation to the potentials of the 
field. After that the value $E$ should be replaced by the operator 
$\varepsilon({\bf\hat p})$.

\section{Interaction of the relativistic spinless particle with a weak 
electromagnetic field}

The problem of description of spinless particle in a weak electromagnetic 
field by one-component wave function with traditional norm condition 
is quite analogous to the description of Dirac particle. That is why it 
is not necessary to explain it in detail and we will write only the main 
results. 

Klein-Gordon equation for a spinless particle in an electromagnetic field 
   \begin{equation} 
   \label{m7}
   (E-e\Phi)^2\psi=(c^2\vec\pi^2+m^2c^4)\psi;
   \end{equation}
can be written in the next form
    \begin{equation}
    \label{m9}
    E\Psi=\left[(\eta+\rho)\frac{{\vec\pi}^2}{2m}+\eta mc^2+e\Phi\right]\Psi;
    \end{equation}
where 
    \begin{equation}   
    \Psi=\left(\begin{array}{cc}\phi\\ \chi \end{array}\right)=\frac12
    \left(\begin{array}{cc}1+(E-U)/mc^2\\1-(E-U)/mc^2\end{array}\right)\psi\,;
    \end{equation}
and $\eta$, $\rho$ are $4\times4$ matrix    
    \begin{equation}
    \eta=\left(\begin{array}{cc}1&0\\0&-1\\ \end{array}\right);\,\,\,\,\,
    \rho=\left(\begin{array}{cc}0&1\\-1&0\\ \end{array}\right).
    \end{equation}
If external electromagnetic field is absent we can separate the "large" and the
"small" components in the equation (\ref{m9}) using the transformation 
$\Psi'=U_0(\hat{\bf\hat p})\Psi$, where   
   \begin{equation}
   U_0({\bf p})=\frac{1}{2\sqrt{mc^2\varepsilon({\bf\hat p})}}
   \left(\varepsilon({\bf\hat p})+mc^2+
   \eta\rho(\varepsilon({\bf\hat p})-mc^2)\right)\,.
   \end{equation}
In the presence of an external electromagnetic field it is not possible to
separate "large" and "small" components in the equation (\ref{m9}) exactly.
In order to make it approximately let us use the operator
$U(\vec\pi)=U_0(\vec\pi)$ for transmission to Foldy-Wouthuysen representation.
Taking into account only values of the zeroth and the first order in small
parameter $e$ we get Shrodinger equation which describe a spinless
particle in a electromagnetic field
   \begin{equation}
   E\psi=\hat H\psi;
   \end{equation}  
where    
   \begin{equation}
   \label{m11}
   \hat H=\varepsilon(\vec\pi)+\frac{e}{2}
   \left(\frac{1}{\sqrt{\varepsilon({\bf\hat p})}}\Phi
   \sqrt{\varepsilon({\bf\hat p})}+
   \sqrt{\varepsilon({\bf\hat p})}\Phi\frac{1}{\sqrt{\varepsilon({\bf\hat p})}}  
   \right).
   \end{equation}
The wave function $\psi$ is one component function and satisfy a traditional
norm condition
   \begin{equation}
   \int\!\psi^+\psi dv=1\,.
   \end{equation}

The Hamiltonian (\ref{m11}) can be expend in series in small parameter $1/c$.
If we took into account several first contributions of this series we get
the relativistic corrections to Hamiltonian of the nonrelativistic spinless
particle \cite{Bjorken}.

\end{document}